\documentclass{optica-article}

\journal{opticajournal}

\articletype{Research Article}

\begin{document}

\title{Measuring the refractive index and thickness of multilayer samples by Fourier domain optical coherence tomography}

\author{Yu-Lin Ku,\authormark{1, 2, *} Yao-Gen Shu\authormark{1, 2, $\dagger$}}

\address{\authormark{1} Wenzhou Institute, University of Chinese Academy of Sciences, Wenzhou, Zhejiang 325000, China\\
\authormark{2} Oujiang Laboratory (Zhejiang Lab for Regenerative Medicine, Vision and Brain Health), Wenzhou, Zhejiang 325024, China\\
}

\email{\authormark{*}kuyulin@ucas.ac.cn\\
\authormark{$\dagger$}shuyaogen@ucas.ac.cn} 


\begin{abstract*} 
Non-contact measurement of the refractive index and thickness of multilayer biological tissues is of great significance for biomedical applications and can greatly improve medical diagnosis and treatment.
In this work, we introduce a theoretical method to simultaneously extract the above information using a Fourier domain optical coherence tomography (FD-OCT) system, in which no additional arrangement and prior information about the object is required other than the OCT interference spectrum.
The single reflection components can be extracted from the observed spectrum by isolating the primary spikes in the sample reflectance profile, and then the refractive index and thickness can be obtained by fitting the actual and modeled values of the single reflection spectrum. 
In a two-layer sample example, the simulation results show that our method can reconstruct the results with high accuracy. 
The relative error is within $\sim 0.01\%$.
The complexity of our approach grows linearly with the number of sample layers, making it well-adapted to multilayer situations.
Our method takes into account both single and multiple reflections in multilayer samples and is therefore equally applicable to samples with high refractive index contrast.

\end{abstract*}

\section{Introduction}

Optical coherence tomography (OCT) is a non-contact imaging technique that generates high-resolution in vivo cross-sectional images without affecting the imaged tissue.
The inception of OCT technology is marked by the development of the time-domain OCT (TD-OCT) system, which was first proposed by Huang et al. and used for in vitro human retinal imaging in 1991~\cite{doi:10.1126/science.1957169}. 
TD-OCT works by utilizing the scanning depth of a reference arm to determine the time of flight of the light signal reflected from the observation sample, resulting in very slow imaging speeds and poor image quality.
Subsequently, the introduction of Fourier-domain OCT (FD-OCT) overcame the limitations of TD-OCT, which was able to capture complete depth information simultaneously~\cite{FERCHER199543}.
In the FD-OCT system, the reference arm is fixed and the detection system is replaced by a spectrometer. 
The Fourier transform of the interference spectrum between the light fields from the reference and the sample arms reveals the internal structure of the object.
FD-OCT has become the preferred method over TD-OCT due to its advantages of significantly improved detection sensitivity and scanning speed, no additional mechanical movement, and direct access to spectral data~\cite{2003Performance, Choma:03, deBoer:03}.

Conventional OCT implementations can only image reflectance profiles that vary with optical path length.
The ability to separate the refractive index and thickness of biological tissue is critical for biomedical applications and can greatly enhance medical diagnosis and treatment.
For example, the refractive index and thickness of the cornea are related to hydration and intraocular pressure status and thus indicate the effects of laser refractive surgery on the cornea~\cite{Kim2004}.
Another important example is the thickness of the human retinal layer, changes of which may be associated with glaucoma~\cite{Mayer10}, diabetes~\cite{Chiu15}, and neuro-ophthalmic diseases including Alzheimer's, Parkinson's, and multiple sclerosis et al~\cite{pmid22677462}.
Accurate measurement of the true thickness and refractive index distribution of the retinal layer can contribute to the early diagnosis and prognosis of related diseases.

Various techniques for simultaneous measurement of refractive index and thickness using OCT systems have been proposed for one-layer~\cite{Sorin124892, Watanabe2014, Cheng10, Chih2014, Wang2006, Tearney:95, Ohmi1997, pmid11008428, Haruna:98, Maruyama:02, Fukano:99, Kim:08, Maruyama2000, Ohmi_2004, Fukano:96,wang2002, Alexandrov:03, Zvyagin:03, Min2012TheRI, Min:13, Hirai:03, MATSUMOTO2006214, Murphy:00, Jin:10, Na:09, Ghim:06, 5770181, 2006Simultaneous} and two- or three-layer~\cite{Lai:14, Zhou:13, 10.1117/1.429972, Tomlins:06, 10.1117/12.645719} objects in both time and Fourier domains.
In Ref.~\cite{10.1117/1.JBO.22.1.015002}, a theoretical method for measuring the refractive index and thickness of samples with an arbitrary number of layers was proposed.
In their approach, the optical path length of each interface was first obtained through the inverse Fourier transform of the FD-OCT interference spectrum, and then the Fresnel coefficient of each interface was obtained through a matrix operation.
The required sample refractive index and thickness can then be extracted directly from the Fresnel equation and the definition of optical path length.
Since the output of their formalism was significantly affected by the uncertainty of the measured optical path length, an optimization method was introduced in their later work to enhance the final results (by sampling the optical path length to achieve the best match between the modeled interference spectrum and the observations)~\cite{10.1117/1.JBO.22.1.015003}.
However, properly selected spectral components of the FD-OCT interference spectrum are required in their method.
Moreover, since they used the summation spectrum to model the interference spectrum, multiple reflections inside the sample were ignored and only single scattering events were considered.
This resulted in their method being able to only handle samples with small refractive index contrast.
To achieve an absolute error $\leq 0.001$, the maximum tolerable refractive index contrast is 0.213.

The primary aim of this study is to present a new theoretical method for the simultaneous measurement of the refractive index and thickness of multilayer systems using FD-OCT.
This article is organized as follows:
In section~\ref{chap:theory}, we give the theoretical framework and methodology for extracting the refractive index and thickness of multilayer samples using the FD-OCT optical spectrum. 
In section~\ref{chap:results}, we demonstrate the feasibility of our approach through numerical simulations in a two-layer system.
We compare the results calculated by our method with their real values and give the discussions.
Finally, we summarize this paper in section~\ref{chap:conclusion}.

\section{Theory}\label{chap:theory}

\begin{figure}[htbp]
\centering\includegraphics[width=\linewidth]{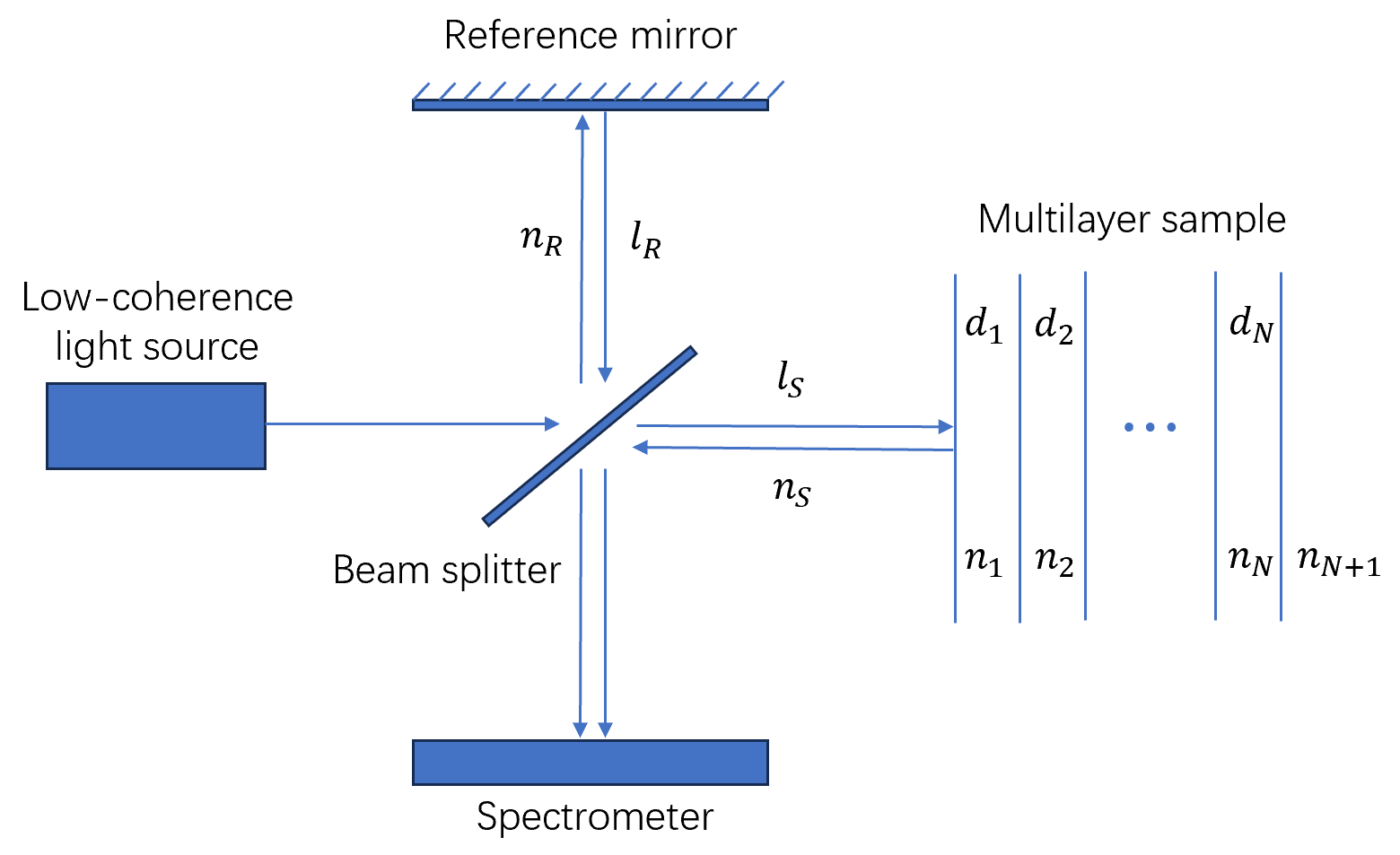}
\caption{Schematic of an FD-OCT system. 
$n_R$ and $n_S$ are the refractive indices of the medium in the reference arm and the sample arm. 
$l_R$ and $l_S$ represent the lengths of the reference arm and the sample arm. 
$(n_i, d_i)$ denotes the refractive index and thickness of each layer of the $N$-layer sample.
$n_{N+1}$ is the refractive index of the medium behind the sample.}
\label{fig:fd_oct}
\end{figure}

An FD-OCT system based on spectral interferometry is depicted in Fig.~\ref{fig:fd_oct}.
A light beam from a low-coherence light source is divided by a beam splitter into two halves.
One half is reflected off the beam splitter and then back-reflected by a reference mirror.
The other half is transmitted through the beam splitter and then back-reflected by the object.
These two back-reflected beams are recombined by the beam splitter and then dispersed by a spectrometer into spectral components.
The corresponding spectral components interfere and form a spectral interferogram acquired by an optical detector array.

The spectral interferogram observed by the detector can be expressed as
\begin{equation}\label{eq:Is}
I(k) = |E_{R}(k) + E_{S}(k)|^{2},
\end{equation}
where $k$ is the wavenumber in vacuum, $E_{R}(k)$ and $E_{S}(k)$ represent the electric field reflected from the reference mirror and the sample, respectively.
The spectral components of the reference and the sample beams have the form of
\begin{equation}\label{eq:E_rs}
\begin{aligned}
&E_{R}(k) = E_{0}(k) r_{R} \exp [i(2kn_Rl_{R} - \omega t)],\\ &
E_{S}(k) = E_{0}(k) r_{S} \exp [i(2kn_Sl_{S} - \omega t)],
\end{aligned}
\end{equation}
where $\omega$ is the angular frequency, $E_{0}(k)$ denotes the electric field incident on the reference mirror and the sample, $r_R$ and $r_S$ stand for the reflectivity of the reference mirror and the sample, $n_R$ and $n_S$ are the refractive indices of the medium in the reference arm and the sample arm, $l_R$ and $l_S$ represent the lengths of the reference arm and the sample arm, respectively. 
Substituting Eq.(\ref{eq:E_rs}) into Eq.(\ref{eq:Is}), we obtain
\begin{equation}\label{eq:Ik}
I(k) = S(k)\left\{ r_R^2+r_S^2+2r_R \mathrm{Re}\left[r_S \exp (2ik\Delta_{RS})\right] \right\}.
\end{equation}
Here, $S(k)$ is the source power spectral density distribution
\begin{equation}
S(k) = |E_0(k)|^2,
\end{equation}
and $\Delta_{RS}$ represents the optical path difference between the sample arm and the reference arm
\begin{equation}
\Delta_{RS} = n_S l_S-n_R l_R.
\end{equation}

The spectral intensity in Eq.(\ref{eq:Ik}) contains three terms.
The first part (proportional to $r_R^2$), referred to as the reference-intensity term, originates from the reference arm reflection.
The second part (proportional to $r_S^2$), referred to as the self-interference term, characterizes the interference among the partial waves from the various sample depths.
The last part, referred to as the cross-interference term, results from the interference between the two light beams from the sample arm and the reference arm.
The first two terms represent the direct current (DC) components while the last term denotes the alternative current (AC) component.
After removing the DC background by phase-shifting interferometry~\cite{Wojtkowski:02}, the normalized spectral response of the AC spectrum has the form of
\begin{equation}\label{eq:I_AC}
\begin{aligned}
I_{\mathrm{AC}}(k) & = \frac{I(k) - I_{\mathrm{D C}}(k)}{2r_RS(k)} \\ &
= \mathrm{Re} \left[r_S \exp (2ik\Delta_{RS})\right].
\end{aligned}
\end{equation}
In traditional FD-OCT, the inverse Fourier transform of the AC spectrum yields an A-line image.

For a $N$-layer sample in Fig.~\ref{fig:fd_oct} which is stratified with ideally flat interfaces and non-absorbing, isotropic, and homogeneous in refractive index, the sample beam consists of multiple partial waves reflecting from each interface.
The AC spectrum can be divided into single and multiple reflection components as
\begin{equation}\label{eq:I_AC_sep}
I_{\mathrm{AC}}(k) = \sum_{j=0}^{N} r_j^{\mathrm{sin}}\cos \left[2k \left(\Delta_{\mathrm{RS}}+\Delta_j^{\mathrm{sin}}\right)\right] + \sum_j^{\mathrm{mul}} r_j^{\mathrm{mul}}\cos \left[2k \left(\Delta_{\mathrm{RS}} + \Delta_j^{\mathrm{mul}}\right)\right],
\end{equation}
where the back reflectance from a single reflection at the $j$-th interface can be determined by the Fresnel formula
\begin{equation}\label{eq:r_sin}
r_j^{\mathrm{sin}} =  \overleftarrow{T}_0 \cdots \overleftarrow{T}_{j-1} R_j \overrightarrow{T}_{j-1} \cdots \overrightarrow{T}_0,
\end{equation}
and the corresponding optical path length is
\begin{equation}\label{eq:od}
\Delta_j^{\mathrm{sin}} = \sum_{l=1}^{j} n_l d_l.
\end{equation}
Here, $R_j$, $\overrightarrow{T}_{j}$ and $\overleftarrow{T}_{j}$ are the Fresnel coefficients of the $j$-th interface
\begin{equation}\label{eq:Fresnel}
\begin{aligned}
& R_j = \frac{n_{j}-n_{j+1}}{n_{j}+n_{j+1}}, \\ &
\overrightarrow{T}_{j} = \frac{2 n_{j}}{n_{j}+n_{j+1}}, \\ &
\overleftarrow{T}_{j} = \frac{2 n_{j+1}}{n_{j}+n_{j+1}}.
\end{aligned}
\end{equation}
$n_l$ and $d_l$ are the refractive index and thickness of the $l$-th sample layer, respectively.
Specifically, at the top of the sample, we have
\begin{equation}
\overrightarrow{T}_{0} = \frac{2n_S}{n_S+n_1},\ \overleftarrow{T}_{0} = \frac{2n_1}{n_S+n_1},\ r_0^{\mathrm{sin}} = \frac{n_S-n_1}{n_S+n_1},\ \Delta_0^{\mathrm{sin}} = 0.
\end{equation}
As for the multiple reflection components, the reflectance scales as $r_j^{\mathrm{mul}}\sim\mathcal{O}(R^j)$ ($j$ stands for the number of reflections and $R$ represents the typical value of Fresnel reflection coefficient), and the corresponding optical path length depends on the actual reflection process.

The profile of the back reflectance versus the optical path length can be obtained by performing an inverse Fourier transform of $I_\mathrm{AC}$.
According to Eq.(\ref{eq:I_AC_sep}), for a multilayer transparent sample, the profile would exhibit a spike-like structure, and the height and position of each spike characterize the reflection intensity and optical path length of the reflection event.
Due to the low refractive index contrast within biological tissue, the reflectance of multiple reflections is much weaker than that of single reflections, making it easy to distinguish between these two reflection cases.
Fourier transforming the reflectance in the neighborhood of each spike back into the wavenumber space can isolate the contribution of each reflection event in the AC spectrum.
We use the Monte Carlo method to search for the reflectance and optical path length in Eq.(\ref{eq:I_AC_sep}) to fit the actual and modeled AC spectrum for each single reflection.
During the fitting process, the fitting function of sum squared residual (SSR) is adopted.
Correctly selecting the search range for the parameters to be fitted can significantly improve the search speed.
We search the optical path length in the neighborhood of each spike to obtain the lowest SSR value.
After determining the reflectance and optical path length of each interface, the refractive index and thickness of each sample layer can be derived from Eq.(\ref{eq:r_sin}), Eq.(\ref{eq:od}) and Eq.(\ref{eq:Fresnel}).

In summary, the process of extracting the refractive index and thickness of each sample layer is as follows:

\begin{enumerate}

\item obtaining the normalized AC spectrum [Eq.(\ref{eq:I_AC})],

\bigskip

\item inverse Fourier transforming the normalized AC spectrum and isolating each primary spike in the reflectance profile,

\bigskip

\item Fourier transforming each primary reflectance spike back into the wavenumber space to obtain the contribution of each single reflection in the AC spectrum,

\bigskip

\item searching the reflectance and optical path length of each sample interface to fit the actual and modeled AC spectrum,

\bigskip

\item calculating the refractive index and thickness of each sample layer through Eq.(\ref{eq:r_sin}), Eq.(\ref{eq:Fresnel}) and Eq.(\ref{eq:od}).

\bigskip

\end{enumerate}

\section{Results and discussions} \label{chap:results}

We verify the feasibility of our method introduced in the previous section with numerically constructed FD-OCT signals through the transfer matrix method (TMM)~\cite{Yeh1988OpticalWI, Knittl1976OpticsOT, OSHeavens_1960}.
The total reflectivity $r_S$ of a multilayer sample can be expressed as
\begin{equation}
r_S = - \frac{Q_{21}}{Q_{22}},
\end{equation}
where 
\begin{equation}
Q = M_{N, N+1}P_{N}M_{N-1, N} \cdots M_{1, 2} P_{1} M_{0, 1}
\end{equation}
is the transfer matrix.
Here, $P_{l}$ and $M_{l, l+1}$ are the propagation and reflection matrices of the $l$-th layer and the $l$-th interface, respectively, which have the form of
\begin{equation}
\begin{gathered}
P_{l} = \left[
        \begin{array}{cc}
        e^{ikn_ld_{l}} & 0 \\
        0 & e^{-ikn_ld_{l}}
        \end{array}
        \right],\\
M_{l, l+1} = \frac{1}{2}
        \left[
        \begin{array}{cc}
        1+n_l/n_{l+1} & 1-n_l/n_{l+1} \\
        1-n_l/n_{l+1} & 1+n_l/n_{l+1}
        \end{array}
        \right].
\end{gathered}
\end{equation}

For simplicity, we consider a two-layer system with two refractive index configurations, one with lower contrast to accommodate the situation in biological tissue: (a) $n_R, n_S, n_3=1.33$, $n_1=1.37$, $n_2=1.45$; and the other with higher contrast: (b) $n_R, n_S, n_3=1$, $n_1=1.4$, $n_2=1.8$.
In both cases, the length difference between the sample arm with the reference arm and the layer thicknesses are set to be $l_S-l_R=200\ \mu m$ and $d_1=300\ \mu m$, $d_2=400\ \mu m$.
The central wavenumber of the light source $k_0$ is taken to be $5.984\ \mu m^{-1}$, corresponding to a central wavelength of $\lambda_0=2\pi/k_0=1050\ nm$, which leads to a relatively low absorption by biological tissues~\cite{Kou:93}.
The spectral bandwidth $W$ is chosen to be $3.142\ \mu m^{-1}$ and the sampling number of wavenumber $P$ is selected as $4000$, which correspond to an axial resolution of $\delta z=\pi/W=1\ \mu m$ and a maximum imaging depth of $z_{\mathrm{max}}=\delta z\cdot P/2=2\ mm$ in vacuum~\cite{Wang2007BiomedicalOP}.
For the two-layer system under consideration, the formula of Eq.(\ref{eq:I_AC_sep}) reduces to
\begin{equation}\label{eq:I_AC_2layer}
\begin{aligned}
I_{\mathrm{AC}}(k) = &r_0\cos (2k\Delta_{RS})+r_1\cos \left[2k\left(\Delta_{RS}+\delta_1\right)\right]+r_2\cos \left[2k\left(\Delta_{RS}+\delta_1+\delta_2\right)\right]\\ &+\mathrm{multiple\ reflection\ terms},
\end{aligned}
\end{equation}
where the reflectances of single reflections are
\begin{equation}\label{eq:r123}
\begin{gathered}
r_0 = \frac{n_0-n_1}{n_0+n_1},\\ 
r_1 = \frac{4n_0n_1}{(n_0+n_1)^2}\frac{n_1-n_2}{n_1+n_2},\\ 
r_2 = \frac{4n_0n_1}{(n_0+n_1)^2}\frac{4n_1n_2}{(n_1+n_2)^2}\frac{n_2-n_3}{n_2+n_3},
\end{gathered}
\end{equation}
and the optical path lengths are
\begin{equation}\label{eq:delta12}
\delta_{1, 2} = n_{1, 2}\cdot d_{1, 2}.
\end{equation}

\begin{figure}[htbp]
\centering\includegraphics[width=\linewidth]{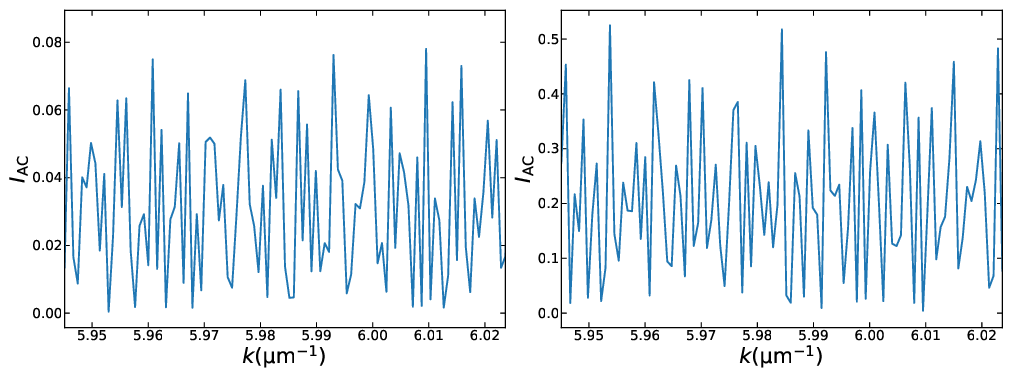}
\caption{The normalized AC spectrum of an FD-OCT system versus wavenumber for a two-layer sample.
The left and right panels correspond to physical parameter configurations (a) and (b) (see text for details), respectively.}
\label{fig:Ik}
\end{figure}

\begin{figure}[htbp]
\centering\includegraphics[width=\linewidth]{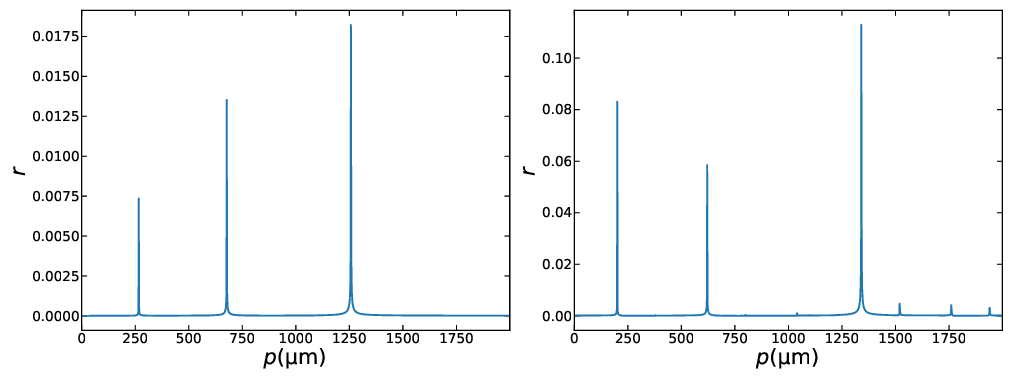}
\caption{Reflectance profile as a function of optical path length, which is derived from the inverse Fourier transform of the normalized AC spectrum $I_{\mathrm{AC}}$.
The left and right panels correspond to physical parameter configurations (a) and (b) (see text for details), respectively.}
\label{fig:rp}
\end{figure}

The normalized spectral response of the AC spectrum calculated from Eq.(\ref{eq:I_AC}) is shown in Fig.~\ref{fig:Ik}, where the left and right panels correspond to the parameter configurations (a) and (b), respectively.
For illustration purposes, only the bandwidth around the central wavenumber is displayed.
Due to the lower refractive index contrast in configuration (a), the intensity of the AC spectrum is much weaker than in configuration (b).
The profile of reflectance with optical path length can be obtained by performing the inverse Fourier transform on the normalized AC spectrum.
The results are shown in Fig.~\ref{fig:rp}.
Similarly, the left panel corresponds to configuration (a), and the right panel corresponds to configuration (b).
As shown in the figure, the reflectance distribution exhibits a spike-like structure, as expected from Eq.(\ref{eq:I_AC_2layer}).
Each spike corresponds to a reflection event, and the individual reflection events are well separated according to their different optical path lengths.
The three primary spikes correspond to single reflections from the three interfaces of the sample, while the much weaker secondary spikes correspond to the results of multiple reflections.
The positions of the three primary spikes agree well with their theoretical predictions.
Due to the much weaker contrast of the refractive index, the spike heights in the left panel are significantly lower than those in the right panel, and the secondary spikes are not obvious.

\begin{figure}[htbp]
\centering\includegraphics[width=\linewidth]{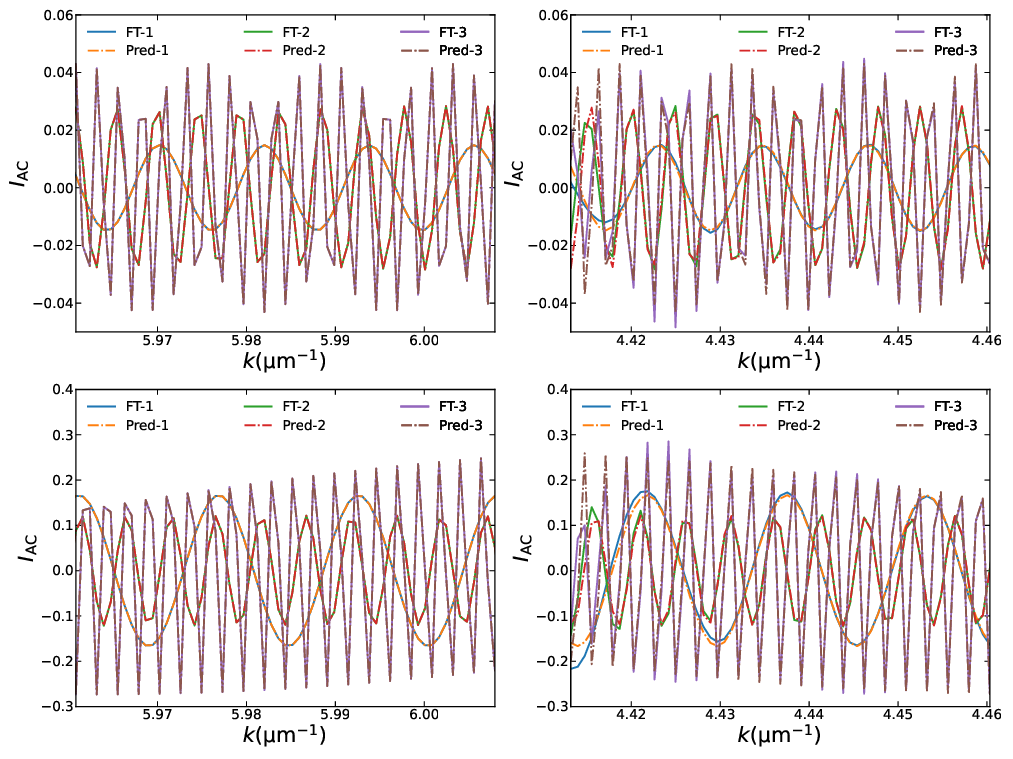}
\caption{AC spectrum for each single reflection versus wavenumber. 
The upper and lower two panels correspond to the physical parameter configurations (a) and (b) (see text for details), respectively.
The left two panels indicate results near the center wavenumber, and the right two panels denote results around the edge wavenumber of the target bandwidth.
The solid lines represent the AC spectrum derived from the Fourier transform of the primary reflectance spikes in Fig.~\ref{fig:rp}, and the dashed lines correspond to the results predicted from Eq.(\ref{eq:I_AC_2layer}).
The numbers indicate the three reflection interfaces of the two-layer sample.
}
\label{fig:Ik_sep}
\end{figure}

By extracting the reflectance values within the neighborhood of each primary spike and Fourier transforming them back into the wavenumber space, the contribution of the single reflection from each sample interface can be separated from the total AC spectrum.
The results are shown in Fig.~\ref{fig:Ik_sep} with solid lines, where the numbers indicate the three reflection interfaces of the sample.
The upper and lower two panels correspond to the physical parameter configurations (a) and (b), respectively.
For presentation purposes, the two panels on the left indicate results within the neighborhood of the center wavenumber, and the two panels on the right denote results around the edge wavenumber of the target bandwidth.
The profile exhibits the expected cosine structure, with the period characterizing the optical path length and the amplitude characterizing the reflectance at the interface.
As a comparison, the AC spectrum for each single reflection predicted from Eq.(\ref{eq:I_AC_2layer}) is displayed with dashed lines in Fig.~\ref{fig:Ik_sep}.
It shows that the AC spectrum values recovered from the reflectance spikes agree well with their modeled values over most of the bandwidth, except for a small portion near the edge wavenumber.

We use the Monte Carlo method to search for the reflectance and optical path length values at each sample interface to fit the actual values of the single reflection spectrum and its modeled values derived from Eq.(\ref{eq:I_AC_2layer}).
The SSR function is adopted to measure the goodness of fit.
We choose the search range of the optical path length to be within the neighborhood of each primary spike location in Fig.~\ref{fig:rp} to speed up the search process.
After obtaining the reflectance and optical path length of each interface, the refractive index and thickness of each sample layer can be derived from Eq.(\ref{eq:r123}) and Eq.(\ref{eq:delta12}).
The results under physical parameter configurations (a) and (b) are shown in Tab.~\ref{tab:nd}. 
In each configuration, the left column is the true physical parameter values, the middle column indicates the deviations between the results recovered by our method with the true values, and the right column corresponds to the relative error (RE) in percentage.
The results show that our method can reconstruct the true physical parameters with high accuracy (relative error within $\sim0. 01\%$), even when the refractive index contrast is large and multiple reflections cannot be ignored.

\begin{table}[htbp]
\centering
\caption{Physical parameters of the two-layer sample considered.
(a) and (b) denote the two parameter configurations (see text for details).
In each configuration, the left column represents the true values, the middle column indicates the deviations between the recovered results and the true values, and the right column corresponds to the RE in percentage.}
\begin{tabular}{c c c c c c c}
\hline \hline
&&a&&&b&\\
\hline
Parameter               & True  & Deviation             & RE(\%)    &       True  & Deviation                       &RE(\%) \\
\hline
$l_S-l_R(\mu m)$        & 200   & $-4.59\times 10^{-5}$ & $-2.29\times 10^{-5}$ & 200   & $+1.53\times 10^{-5}$ & $+7.63\times 10^{-6}$  \\

$d_1(\mu m)$            & 300   & $2.17\times 10^{-3}$  & $+7.23\times 10^{-4}$ & 300   & $+2.94\times 10^{-2}$ & $+9.8\times 10^{-3}$   \\

$d_2(\mu m)$            & 400   & $-9.84\times 10^{-3}$ & $-2.46\times 10^{-3}$ & 400   & $-4.65\times 10^{-2}$ & $-1.16\times 10^{-2}$   \\

$n_1$                   & 1.33  & $+5.75\times 10^{-5}$ & $+4.32\times 10^{-3}$ & 1     & $+1.11\times 10^{-4}$ & $+1.11\times 10^{-2}$   \\

$n_2$                   & 1.37  & $-9.7\times 10^{-6}$  & $-7.08\times 10^{-4}$ & 1.4   & $-1.37\times 10^{-4}$ & $-9.8\times 10^{-3}$   \\

$n_3$                   & 1.45  & $+3.57\times 10^{-5}$ & $+2.46\times 10^{-3}$ & 1.8   & $+2.09\times 10^{-4}$ & $+1.16\times 10^{-2}$   \\
\hline \hline
\end{tabular}
\label{tab:nd}
\end{table}

\section{Conclusion}\label{chap:conclusion}

In summary, we propose a new theoretical framework in this article to extract the refractive index and thickness of a multilayer sample simultaneously through an FD-OCT system, without any prior information about the sample required.
Our method works as follows:
First, the sample reflectance profile versus optical path length is obtained by performing an inverse Fourier transform of the observed AC spectrum;
Second, we separate the primary spikes in the reflectance profile;
Then, each single reflection component in the total AC spectrum is extracted by Fourier transforming the primary reflectance spikes back into wavenumber space;
Finally, we use the Monte Carlo method to search for the target physical parameters to fit the actual and modeled AC spectrum from each single reflection.
Taking a two-layer system as an example, we verify the feasibility of our approach.
The results show that our method can reconstruct the true sample refractive index and thickness with high accuracy.
The relative error is within $\sim 0.01\%$.
Extending to the cases of more layers ($N$) is straightforward.
Unlike methods that search for all physical parameters simultaneously with a complexity of $\mathcal{O}(\mathbb{R}^N)$, the complexity of our method scales as $\mathcal{O}(N\cdot \mathbb{R})$.
This allows our method to be better adapted to the cases of multilayer samples.
In addition, our theoretical framework takes into account not only single but also multiple reflection events, thus eliminating the need for samples to have low refractive index contrast.
When the reflectivity varies significantly among different interfaces, it will be difficult to distinguish single and multiple reflections from the reflectance profile.
In this case, our approach would no longer be applicable and additional imaging methods can be employed to determine the interface locations.

\begin{backmatter}

\bmsection{Funding}
Wenzhou Institute of UCAS (WIUCASQD2020009, WIUCASSICTP2022), Oujiang Laboratory (Zhejiang Lab for Regenerative Medicine, Vision and Brain Health) (OJQDJQ2022001).

\bmsection{Acknowledgments}
The Discipline Cluster for Oncology of Wenzhou Medical University under the grant z1-2023005.

\bmsection{Disclosures}
The authors declare that there are no conflicts of interest related to this article.

\end{backmatter}

\bibliography{References}

\begin{thebibliography}{10}
\newcommand{\enquote}[1]{``#1''}

\bibitem{doi:10.1126/science.1957169}
D.~Huang, E.~A. Swanson, C.~P. Lin, \emph{et~al.}, \enquote{Optical coherence tomography,} {\protect\JournalTitle{Science}} \textbf{254}, 1178--1181 (1991).

\bibitem{FERCHER199543}
A.~Fercher, C.~Hitzenberger, G.~Kamp, and S.~El-Zaiat, \enquote{Measurement of intraocular distances by backscattering spectral interferometry,} {\protect\JournalTitle{Optics Communications}} \textbf{117}, 43--48 (1995).

\bibitem{2003Performance}
R.~Leitgeb, C.~K. Hitzenberger, and A.~F. Fercher, \enquote{Performance of fourier domain vs. time domain optical coherence tomography,} {\protect\JournalTitle{Optics Express}} \textbf{11}, 889--894 (2003).

\bibitem{Choma:03}
M.~A. Choma, M.~V. Sarunic, C.~Yang, and J.~A. Izatt, \enquote{Sensitivity advantage of swept source and fourier domain optical coherence tomography,} {\protect\JournalTitle{Opt. Express}} \textbf{11}, 2183--2189 (2003).

\bibitem{deBoer:03}
J.~F. de~Boer, B.~Cense, B.~H. Park, \emph{et~al.}, \enquote{Improved signal-to-noise ratio in spectral-domain compared with time-domain optical coherence tomography,} {\protect\JournalTitle{Opt. Lett.}} \textbf{28}, 2067--2069 (2003).

\bibitem{Kim2004}
Y.~L. Kim, J.~T. Walsh, T.~K. Goldstick, and M.~R. Glucksberg, \enquote{Variation of corneal refractive index with hydration,} {\protect\JournalTitle{Physics in Medicine and Biology}} \textbf{49}, 859 (2004).

\bibitem{Mayer10}
M.~A. Mayer, J.~Hornegger, C.~Y. Mardin, and R.~P. Tornow, \enquote{Retinal nerve fiber layer segmentation on fd-oct scans of normal subjects and glaucoma patients,} {\protect\JournalTitle{Biomed. Opt. Express}} \textbf{1}, 1358--1383 (2010).

\bibitem{Chiu15}
S.~J. Chiu, M.~J. Allingham, P.~S. Mettu, \emph{et~al.}, \enquote{Kernel regression based segmentation of optical coherence tomography images with diabetic macular edema,} {\protect\JournalTitle{Biomed. Opt. Express}} \textbf{6}, 1172--1194 (2015).

\bibitem{pmid22677462}
T.~Garcia, A.~Tourbah, E.~Setrouk, \emph{et~al.}, \enquote{Optical coherence tomography in neuro-ophthalmology,} {\protect\JournalTitle{J Fr Ophtalmol}} \textbf{35}, 454--466 (2012).

\bibitem{Sorin124892}
W.~Sorin and D.~Gray, \enquote{Simultaneous thickness and group index measurement using optical low-coherence reflectometry,} {\protect\JournalTitle{IEEE Photonics Technology Letters}} \textbf{4}, 105--107 (1992).

\bibitem{Watanabe2014}
K.~Watanabe, M.~Ohshima, and T.~Nomura, \enquote{Simultaneous measurement of refractive index and thickness distributions using low-coherence digital holography and vertical scanning,} {\protect\JournalTitle{Journal of Optics}} \textbf{16}, 045403 (2014).

\bibitem{Cheng10}
H.-C. Cheng and Y.-C. Liu, \enquote{Simultaneous measurement of group refractive index and thickness of optical samples using optical coherence tomography,} {\protect\JournalTitle{Appl. Opt.}} \textbf{49}, 790--797 (2010).

\bibitem{Chih2014}
C.-T. Yen, J.-F. Huang, M.-J. Wu, \emph{et~al.}, \enquote{{Simultaneously measuring the refractive index and thickness of an optical sample by using improved fiber-based optical coherence tomography},} {\protect\JournalTitle{Optical Engineering}} \textbf{53}, 044108 (2014).

\bibitem{Wang2006}
Y.-P.~W. et~al., \enquote{Reflectometry measuring refractive index and thickness of polymer samples simultaneously,} {\protect\JournalTitle{Journal of Modern Optics}} \textbf{53}, 1845--1851 (2006).

\bibitem{Tearney:95}
G.~J. Tearney, M.~E. Brezinski, J.~F. Southern, \emph{et~al.}, \enquote{Determination of the refractive index of highly scattering human tissue by optical coherence tomography,} {\protect\JournalTitle{Opt. Lett.}} \textbf{20}, 2258--2260 (1995).

\bibitem{Ohmi1997}
M.~Ohmi, T.~Shiraishi, H.~Tajiri, and M.~Haruna, \enquote{Simultaneous measurement of refractive index and thickness of transparent plates by low coherence interferometry,} {\protect\JournalTitle{Optical Review}} \textbf{4}, 507--515 (1997).

\bibitem{pmid11008428}
M.~Ohmi, Y.~Ohnishi, K.~Yoden, and M.~Haruna, \enquote{{{I}n vitro simultaneous measurement of refractive index and thickness of biological tissue by the low coherence interferometry},} {\protect\JournalTitle{IEEE Trans Biomed Eng}} \textbf{47}, 1266--1270 (2000). [DOI:\href{https://dx.doi.org/10.1109/10.867961}{10.1109/10.867961}] [PubMed:\href{https://www.ncbi.nlm.nih.gov/pubmed/11008428}{11008428}].

\bibitem{Haruna:98}
M.~Haruna, M.~Ohmi, T.~Mitsuyama, \emph{et~al.}, \enquote{Simultaneous measurement of the phase and group indices and the thickness of transparent plates by low-coherence interferometry,} {\protect\JournalTitle{Opt. Lett.}} \textbf{23}, 966--968 (1998).

\bibitem{Maruyama:02}
H.~Maruyama, S.~Inoue, T.~Mitsuyama, \emph{et~al.}, \enquote{Low-coherence interferometer system for the simultaneous measurement of refractive index and thickness,} {\protect\JournalTitle{Appl. Opt.}} \textbf{41}, 1315--1322 (2002).

\bibitem{Fukano:99}
T.~Fukano and I.~Yamaguchi, \enquote{Separation of measurement of the refractive index and the geometrical thickness by use of a wavelength-scanning interferometer with a confocal microscope,} {\protect\JournalTitle{Appl. Opt.}} \textbf{38}, 4065--4073 (1999).

\bibitem{Kim:08}
S.~Kim, J.~Na, M.~J. Kim, and B.~H. Lee, \enquote{Simultaneous measurement of refractive index and thickness by combining low-coherence interferometry and confocal optics,} {\protect\JournalTitle{Opt. Express}} \textbf{16}, 5516--5526 (2008).

\bibitem{Maruyama2000}
H.~Maruyama, T.~Mitsuyama, M.~Ohmi, and M.~Haruna, \enquote{Simultaneous measurement of refractive index and thickness by low coherence interferometry considering chromatic dispersion of index,} {\protect\JournalTitle{Optical Review}} \textbf{7}, 468--472 (2000).

\bibitem{Ohmi_2004}
M.~Ohmi, H.~Nishi, Y.~Konishi, \emph{et~al.}, \enquote{High-speed simultaneous measurement of refractive index and thickness of transparent plates by low-coherence interferometry and confocal optics,} {\protect\JournalTitle{Measurement Science and Technology}} \textbf{15}, 1531 (2004).

\bibitem{Fukano:96}
T.~Fukano and I.~Yamaguchi, \enquote{Simultaneous measurement of thicknesses and refractive indices of multiple layers by a low-coherence confocal interference microscope,} {\protect\JournalTitle{Opt. Lett.}} \textbf{21}, 1942--1944 (1996).

\bibitem{wang2002}
X.~Wang, C.~Zhang, L.~Zhang, \emph{et~al.}, \enquote{{Simultaneous refractive index and thickness measurements of bio-tissue by optical coherence tomography},} {\protect\JournalTitle{Journal of Biomedical Optics}} \textbf{7}, 628 -- 632 (2002).

\bibitem{Alexandrov:03}
S.~A. Alexandrov, A.~V. Zvyagin, K.~K. M. B.~D. Silva, and D.~D. Sampson, \enquote{Bifocal optical coherenc refractometry of turbid media,} {\protect\JournalTitle{Opt. Lett.}} \textbf{28}, 117--119 (2003).

\bibitem{Zvyagin:03}
A.~V. Zvyagin, K.~K. M. B.~D. Silva, S.~A. Alexandrov, \emph{et~al.}, \enquote{Refractive index tomography of turbid media by bifocal optical coherence refractometry,} {\protect\JournalTitle{Opt. Express}} \textbf{11}, 3503--3517 (2003).

\bibitem{Min2012TheRI}
G.~Min, J.~W. Kim, and B.~H. Lee, \enquote{The refractive index measurement technique based on the defocus correction method in full-field optical coherence tomography,} in \emph{Photonics Europe,}  (2012).

\bibitem{Min:13}
G.~Min, W.~J. Choi, J.~W. Kim, and B.~H. Lee, \enquote{Refractive index measurements of multiple layers using numerical refocusing in ff-oct,} {\protect\JournalTitle{Opt. Express}} \textbf{21}, 29955--29967 (2013).

\bibitem{Hirai:03}
A.~Hirai and H.~Matsumoto, \enquote{Low-coherence tandem interferometer for measurement of group refractive index without knowledge of the thickness of the test sample,} {\protect\JournalTitle{Opt. Lett.}} \textbf{28}, 2112--2114 (2003).

\bibitem{MATSUMOTO2006214}
H.~Matsumoto, K.~Sasaki, and A.~Hirai, \enquote{In situ measurement of group refractive index using tandem low-coherence interferometer,} {\protect\JournalTitle{Optics Communications}} \textbf{266}, 214--217 (2006).

\bibitem{Murphy:00}
D.~F. Murphy and D.~A. Flavin, \enquote{Dispersion-insensitive measurement of thickness and group refractive index by low-coherence interferometry,} {\protect\JournalTitle{Appl. Opt.}} \textbf{39}, 4607--4615 (2000).

\bibitem{Jin:10}
J.~Jin, J.~W. Kim, C.-S. Kang, \emph{et~al.}, \enquote{Thickness and refractive index measurement of a silicon wafer based on an optical comb,} {\protect\JournalTitle{Opt. Express}} \textbf{18}, 18339--18346 (2010).

\bibitem{Na:09}
J.~Na, H.~Y. Choi, E.~S. Choi, \emph{et~al.}, \enquote{Self-referenced spectral interferometry for simultaneous measurements of thickness and refractive index,} {\protect\JournalTitle{Appl. Opt.}} \textbf{48}, 2461--2467 (2009).

\bibitem{Ghim:06}
Y.-S. Ghim and S.-W. Kim, \enquote{Thin-film thickness profile and its refractive index measurements by dispersive white-light interferometry,} {\protect\JournalTitle{Opt. Express}} \textbf{14}, 11885--11891 (2006).

\bibitem{5770181}
S.~J. Park, K.~S. Park, Y.~H. Kim, and B.~H. Lee, \enquote{Simultaneous measurements of refractive index and thickness by spectral-domain low coherence interferometry having dual sample probes,} {\protect\JournalTitle{IEEE Photonics Technology Letters}} \textbf{23}, 1076--1078 (2011).

\bibitem{2006Simultaneous}
P.~H. Tomlins and R.~K. Wang, \enquote{Simultaneous analysis of refractive index and physical thickness by fourier domain optical coherence tomography,} {\protect\JournalTitle{Optoelectronics Iee Proceedings}} \textbf{153}, 222--228 (2006).

\bibitem{Lai:14}
T.~Lai and S.~Tang, \enquote{Cornea characterization using a combined multiphoton microscopy and optical coherence tomography system,} {\protect\JournalTitle{Biomed. Opt. Express}} \textbf{5}, 1494--1511 (2014).

\bibitem{Zhou:13}
Y.~Zhou, K.~K.~H. Chan, T.~Lai, and S.~Tang, \enquote{Characterizing refractive index and thickness of biological tissues using combined multiphoton microscopy and optical coherence tomography,} {\protect\JournalTitle{Biomed. Opt. Express}} \textbf{4}, 38--50 (2013).

\bibitem{10.1117/1.429972}
A.~R. Knuettel and M.~Boehlau-Godau, \enquote{{Spatially confined and temporally resolved refractive index and scattering evaluation in human skin performed with optical coherence tomography},} {\protect\JournalTitle{Journal of Biomedical Optics}} \textbf{5}, 83 -- 92 (2000).

\bibitem{Tomlins:06}
P.~H. Tomlins and R.~K. Wang, \enquote{Matrix approach to quantitative refractive index analysis by fourier domain optical coherence tomography,} {\protect\JournalTitle{J. Opt. Soc. Am. A}} \textbf{23}, 1897--1907 (2006).

\bibitem{10.1117/12.645719}
P.~H. Tomlins and R.~K. Wang, \enquote{{Layer dependent refractive index measurement by Fourier domain optical coherence tomography},} International Society for Optics and Photonics (SPIE, 2006), p. 607913.

\bibitem{10.1117/1.JBO.22.1.015002}
P.~Rajai, H.~Schriemer, A.~Amjadi, and R.~Munger, \enquote{{Simultaneous measurement of refractive index and thickness of multilayer systems using Fourier domain optical coherence tomography, part 1: theory},} {\protect\JournalTitle{Journal of Biomedical Optics}} \textbf{22}, 015002 (2017).

\bibitem{10.1117/1.JBO.22.1.015003}
P.~Rajai, H.~Schriemer, A.~Amjadi, and R.~Munger, \enquote{{Simultaneous measurement of refractive index and thickness of multilayer systems using Fourier domain optical coherence tomography, part 2: implementation},} {\protect\JournalTitle{Journal of Biomedical Optics}} \textbf{22}, 015003 (2017).

\bibitem{Wojtkowski:02}
M.~Wojtkowski, A.~Kowalczyk, R.~Leitgeb, and A.~F. Fercher, \enquote{Full range complex spectral optical coherence tomography technique in eye imaging,} {\protect\JournalTitle{Opt. Lett.}} \textbf{27}, 1415--1417 (2002).

\bibitem{Yeh1988OpticalWI}
P.~A. Yeh and M.~T. Hendry, \enquote{Optical waves in layered media,}  (1988).

\bibitem{Knittl1976OpticsOT}
Z.~Knittl and T.~Z. Knittl., \enquote{Optics of thin films; an optical multilayer theory,}  (1976).

\bibitem{OSHeavens_1960}
O.~S. Heavens, \enquote{Optical properties of thin films,} {\protect\JournalTitle{Reports on Progress in Physics}} \textbf{23}, 1 (1960).

\bibitem{Kou:93}
L.~Kou, D.~Labrie, and P.~Chylek, \enquote{Refractive indices of water and ice in the 0.65- to 2.5-$\mu$m spectral range,} {\protect\JournalTitle{Appl. Opt.}} \textbf{32}, 3531--3540 (1993).

\bibitem{Wang2007BiomedicalOP}
L.~V. Wang and H.~Wu, \enquote{Biomedical optics: Principles and imaging,}  (2007).

\end{thebibliography}

\end{document}